# Avoiding Negative Probabilities in Quantum Mechanics

**Golden Gadzirayi Nyambuya**

Department of Applied Physics, National University of Science and Technology, Bulawayo, Republic of Zimbabwe
Email: physicist.ggn@gmail.com

## ABSTRACT

As currently understood since its discovery, the bare Klein-Gordon theory consists of negative quantum probabilities which are considered to be physically meaningless if not outright obsolete. Despite this annoying setback, these negative probabilities are what led the great Paul Dirac in 1928 to the esoteric discovery of the Dirac Equation. The Dirac Equation led to one of the greatest advances in our understanding of the physical world. In this reading, we ask the seemingly senseless question, "Do negative probabilities exist in quantum mechanics?" In an effort to answer this question, we arrive at the conclusion that depending on the choice one makes of the quantum probability current, one will obtain negative probabilities. We thus propose a new quantum probability current of the Klein-Gordon theory. This quantum probability current leads directly to positive definite quantum probabilities. Because these negative probabilities are in the bare Klein-Gordon theory, intrinsically a result of negative energies, the fact that we here arrive at a theory with positive probabilities, means that negative energy particles are not to be considered problematic as is the case in the bare Klein-Gordon theory. From an abstract—objective stand-point; in comparison with positive energy particles, the corollary is that negative energy particles should have equal chances to exist. As to why these negative energy particles do not exist, this is analogous to asking why is it that Dirac's antimatter does not exist in equal proportions with matter. This problem of why negative energy particles do not exist in equal proportions with positive energy particles is a problem that needs to be solved by a future theory.

## 1. Introduction

*"My work always tried to unite the Truth with the Beautiful, but when I had to choose one or the other, I usually chose the Beautiful."*

—Hermann Klaus Hugo Weyl (1885-1955)

If one accepts the bare Klein-Gordon theory as it is currently understood since its discovery in 1927 by Oskar Klein (1894-1977, of Sweden) and Walter Gordon (1893-1939, of Germany), then, there is no doubt that they will accept without fail that negative quantum mechanical probabilities do exist in the bare Klein-Gordon theory. Solemnly, by a combination of a deep and rare curiosity, fortune, and serendipity, than by natural design, the existence of these negative probabilities in the Klein-Gordon theory is what led the eminent British physicist Paul Adrien Maurice Dirac (1902-1984) to his landmark discovery of the Dirac Equation [1,2]. Needless to say, but perhaps as a way of expressing our deepest admiration of this great achievement, the Dirac Equation ranks amongst the greatest and most noble equations of physics. Eighty four years on since its discovery (*i.e.*, 1928-2012), the Dirac Equation is an equation whose wealth of knowledge cannot be said to have been completely deciphered and fathomed but is in the process thereof.

Like other deep-thinking physicists of his time, right from the-word-go, Dirac passionately objected to the notion of negative probabilities implied by the Klein-Gordon theory. This led him to silently embark on a noble scientific journey of the mind whose final destination was to successfully solve this persistent and nagging problem of negative probability. Dirac had hoped that by eliminating the negative probabilities, he would concurrently eliminate the negative energies—he hoped for the rare fortuity of hitting two birds with one stone. Alas, that did not happen. Only the possibility of negative probabilities vanished while the negative energies stubbornly reared their "ugly head" in Dirac's new vision. On completion of his seemingly divine scientific journey of the mind, he arrived at his esoteric equation, which accurately describes the Electron.

Why do we say Dirac silently embarked on his quest for the Dirac Equation? Well the answer to this is that—for example, during a (tea/lunch) break at the 1927 Solvay conference attended by the great Danish physicist Neils Henrik David Bohr (1885-1962), Dirac, and many other towering figures of the last half century in physics; Dirac was asked by Neils Bohr what he was working on, to which he replied: "I'm trying to take the square root of something..." meaning the square root of the Klein-Gordon Equation—in physics, this is strange because it meant taking the square root of an operator. Later, Dirac recalled that he continued on by saying he was trying to find a relativistic quantum theory of the Electron, to which Bohr commented, "But Klein has already solved that problem". Dirac then tried to explain he was not satisfied with the (Klein-Gordon) solution because it involved a $2^{nd}$ order equation in the time and space derivatives. Dirac was simply not open; he was a man of notoriously very few words, he meant every word he said; in a nutshell, he was economic with his words.

Further, Dirac was a man of great mathematical subtleness; it is this quality which led him to the Dirac Equation. He believed that one must follow the mathematics to where it would lead and in so doing he unlocked a great wealth of ideas such as magnetic monopoles, the variation of the gravitational constant amongst others. On the same pedestal—in 1942, all in an effort to forge ahead to seek new frontiers in the field of physical knowledge; Dirac back-pedalled on negative probabilities when he delved once again into the then un-chartered waters of negative probability when he wrote a paper entitled: "The Physical Interpretation of Quantum Mechanics" where he introduced the concept of negative probabilities [3]. In the introductory section of his watershed reading, he said:

"*Negative energies and probabilities should not be considered as* [*mere*] *nonsense. They are well-defined concepts mathematically, like a negative of money*".

Fifty five years later after Dirac's musings, *i.e.* in 1987 toward the end of his fruitful life, another great mind, the flamboyant and charismatic American physicist, Richard Feynman (1918-1988), took the idea further when he argued that, no one objects to using negative numbers in calculations, although "minus three apples" is not a valid concept in real life [so, it should be reasonable to consider negative probabilities too]. Further into the shores of the unknown, he [Feynman] argued not only how negative probabilities could possibly be useful in probability calculations, but as well how probabilities above unity may be useful [4]. Really? What would a probability above unity really mean?

The ideas of Dirac and Feynman on negative probability have not gained much support. To ourselves, negative probabilities, even if they may be well defined mathematical concepts as Dirac and Feynman believe or want us to believe, they [negative probabilities] are physically meaningless and obsolete; they signify something "sinister" in the finer detail of the theory in question. We like to view these ideas of negative probability as nothing more than highlighting and dramatising the desperation by physicists to make sense of nonsense all in an effort to find a natural explanation of nonsense. We think and hold that nonsense is nonsense and should be left that way; one should simply let the sleeping dogs lay. We have to be ruthless in our dismissal of these negative probabilities; we can only hope that the reader will pardon us for this.

The root of negative probabilities and all that seems bizarre in the world of probability theory is the Klein-Gordon theory. If it could be shown that the Klein-Gordon theory is devoid of these, it would render Dirac and Feynman's effort worthless—and; order, harmony and tranquillity will certainly be restored in the world of probability theory. This would mean the chapter of negative probabilities is closed altogether. The endeavour of this reading is to point out that the Klein-Gordon theory is devoid of these negative probabilities, hence possibly bringing to a complete standstill Dirac and Feynman's efforts.

If only physicists had extended the British-German physicist Max Born (1882-1970)'s idea that the magnitude of the wavefunction gives the probability density function [7]; that is, extend this idea so that it applies to all quantum mechanical wavefunctions, then, we would never have landed on these rough, bizarre and uncertain shores of negative probabilities. As will be argued, what physicists have done is to carry over the probability current density found in the Schrödinger theory directly into the Klein-Gordon theory, in which process the quantum probability of the Klein-Gordon theory is constrained in a manner that allows for negative probabilities. Our suggestion, if correct as we would like to believe, is that instead of carrying over the probability current density found in the Schrödinger theory into the Klein-Gordon theory, we need to do things the other way round, that is, we have to carry over the probability density function of the Schrödinger theory into the Klein-Gordon theory. Simple, we must generalize Born's idea, that is:

*Born's idea that the wavefunction represents the probability amplitude and its magnitude represents the probability*; *this idea must be generalized so that it is applicable to any general wavefunction that purports to describe undulatory material particles.*

In this way, we constrain the resultant probability current density of the Klein-Gordon theory so as to allow only for positive definite probabilities—this is where we believe the problem in the negative probabilities lies. In the end, we obtain a Klein-Gordon theory that is devoid of negative probabilities; this of cause leads us to an objective world since all the probabilities are not only positive but positive definite. Notice that Schrödinger's wavefunction together with Dirac's wavefunction all conform to Born's idea but the Klein-Gordon wavefunction does not. Why? We herein propose an answer to this question.

Now, to wind up this section, we shall give the synopsis of this reading; it is organised as follows: in the next section, we present the Schrödinger quantum mechanical probability theory as it is understood in the present day. In §(3), we also present the Klein-Gordon probability theory as it is understood in the present day. In §(4), we go onto the main theme of the present reading, where we demonstrate that if one makes an appropriate choice of the Klein Gordon probability current, one obtains a Klein-Gordon theory that is free from negative probabilities. While §(2) and §(3) may seem trivial to the quantum mechanically erudite reader, it is worthwhile that we mention that we have taken the decision to go through these sections [*i.e.* §(2) and §(3)] for nothing other than instructive purposes. The well versed reader will obviously have to skip these and go straight to §(4). In §(5), we give the overall discussion and conclusions drawn thereof.

## 2. Schrödinger Theory

While in search of the Schrödinger Equation the great Austrian physicist, Erwin Rudolf Josef Alexander Schrödinger (1887-1961) first arrived at the Klein-Gordon Equation but discarded it because it did not give the correct predictions for the hydrogen atom. Schrödinger, was largely motivated to successfully search for the Schrödinger Equation after a thoughtful remark by the eminent Professor, Peter Joseph William Debye (1884-1966, of Austria), at the end of a lecture that he [Schrödinger] delivered on *de* Broglie's waves at the University of Vienna where he [Schrödinger] was working.

Professor Debye who was the head of the physics research group, on hearing of the *de* Broglie waves, he asked Schrödinger to explain these to the rest of the research group. So, the great Schrödinger weighed up to the task. At the end of the lecture, Professor Debye remarked that it seemed childish to talk of waves without a corresponding wave equation?! This proved to be Schrödinger's great moment of inspiration that would immortalize his name in the annals of human history. What is the wave equation describing the *de* Broglie waves?

In his 1924 doctoral thesis, which was nearly turned down [thanks to the pre-eminent French physicist Paul Langevin (1872-1946)'s wisdom and Einstein's influence and stature[1]], the French Prince, Louis Victor Pierre Raymond *de* Broglie (1892-1987) hypothesised that there is a duality between waves and matter; he gave a formula for the matter waves which stated that the wavelength of material particles is inversely proportional to the momentum of the matter particle in question. However, in his proposal, he did not propose the corresponding wave equation for these matter waves. As Professor Debye pointed out to Schrödinger, logic dictates that every wave must be described by a corresponding wave equation. The deep-and-agile Schrödinger saw immediately the depth and breadth of Professor Debye's question and as the lore holds; he [Schrödinger] went into "hiding" for about six months in search of the Schrödinger Equation which he successfully found at the end of his esoteric sojourn which was not without tribulation and trials [6]. The equation he found is:

$$-\frac{\hbar^2}{2m}\nabla^2\Psi + V\Psi = i\hbar\frac{\partial\Psi}{\partial t}, \qquad (1)$$

where the arcane symbol $\Psi$ is the Schrödinger wavefunction (or probability amplitude), $\hbar$ is Planck's normalized constant, $m$ is the mass of the particle in question and $\nabla$ is the usual $3D$ differential operator and $t$ is time. In presenting his equation in 1926, Schrödinger interpreted the magnitude of the wavefunction as giving the density of the Electronic charge of the atom. However, Max Born [3] gave a radically different interpretation, where this quantity [magnitude of the wavefunction] is assumed to represent the probability that the atom is in a given state. In this way, Born ushered physics into the depth and realm of probability calculus. To this day, physicists do not agree on how to interpret this wavefunction but the general consensus is that, it is a probability function.

### 2.1. Schrödinger Probability Current Density

For instructive purposes, we present here the usual way in which one arrives at the expression of the Schrödinger probability current density. To do this, we have to take the Schrödinger Equation, divide it throughout by $i\hbar$ and then multiply the resultant by the complex conjugate of the wavefunction, *i.e.*:

---

[1]Instead of rejecting *de* Broglie's doctoral thesis, the cautious Langevin decided that the greatest living scientist of the day—Albert Einstein—must be consulted on the matter before a fateful decision is reached. The agile Einstein immediately endorsed the idea with the remark that *de* Broglie had gone a step further from where he [Einstein] had left the issue of wave-particle duality. In his [Einstein] own words: "I believe it is a first feeble *ray of light on this worst of our physics enigmas. I, too, have found something which speaks for his c*onstruction" [see e.g. Ref. 5, p. 21]. Behold! To a "sleepy world", unknowingly, like Russia's Sputnik flying past the United States of America in the cover of darkness, *de* Broglie had just opened the scientific Pandora's box, a very rich scientific gold mine which to the present we are still trying to understand.

$$\Psi^* \frac{\partial \Psi}{\partial t} = \frac{i\hbar}{2m} \left( \Psi^* \nabla^2 \Psi - \frac{2imV}{\hbar} \Psi^* \Psi \right). \quad (2)$$

Further, taking the complex conjugate of this same equation and then multiplying it by the wavefunction, one arrives at:

$$\Psi \frac{\partial \Psi^*}{\partial t} = -\frac{i\hbar}{2m} \left( \Psi \nabla^2 \Psi^* - \frac{2imV}{\hbar} \Psi^* \Psi \right), \quad (3)$$

and now adding these two equations, one obtains:

$$\Psi^* \frac{\partial \Psi}{\partial t} + \Psi \frac{\partial \Psi^*}{\partial t} = \frac{i\hbar}{2m} \left( \Psi^* \nabla^2 \Psi - \Psi \nabla^2 \Psi^* \right). \quad (4)$$

If we set $\rho = \Psi^* \Psi$, which is the probability density function, then:

$$\frac{\partial \rho}{\partial t} = \Psi^* \frac{\partial \Psi}{\partial t} + \Psi \frac{\partial \Psi^*}{\partial t}. \quad (5)$$

For the right handside of Equation (4), we will have:

$$\frac{i\hbar}{2m} \left( \Psi^* \nabla^2 \Psi - \Psi \nabla^2 \Psi^* \right)
= -\nabla \cdot \left[ \frac{i\hbar}{2m} \left( \Psi^* \nabla \Psi - \Psi \nabla \Psi^* \right) \right], \quad (6)$$

and setting:

$$\boldsymbol{J}_\rho^S = \frac{i\hbar}{2m} \left( \Psi^* \nabla \Psi - \Psi \nabla \Psi^* \right), \quad (7)$$

which is the Schrödinger probability current density, then, Equation (4) reduces the probability conservation equation, *i.e.*:

$$\frac{\partial \rho}{\partial t} = -\nabla \cdot \boldsymbol{J}_\rho^S. \quad (8)$$

Now, this is not the only continuity equation that can be written out of the Schrödinger Equation. The quantity $\left( \rho, \boldsymbol{J}_\rho^S \right)$ comprises the four probability current density. In the next section, we shall write another continuity equation out of the Schrödinger Equation not in terms of the probability density function but in terms of the probability amplitude, $\Psi$.

## 2.2. Schrödinger Equation as a Continuity Equation

To write down another continuity equation out of the Schrödinger Equation in terms of the probability amplitude, we know that, for any general smooth, square integrable and differentiable space and time varying function (or field) $\Psi = \Psi(\boldsymbol{r},t)$, from the chain rule, the following holds true always:

$$\Psi \equiv \int_{r_0}^{r} (\nabla \Psi) \cdot d\boldsymbol{r} = \int_{t_0}^{t} (\nabla \Psi) \cdot \left( \frac{d\boldsymbol{r}}{dt} \right) dt
= \int_{t_0}^{t} (\boldsymbol{v} \cdot \nabla \Psi) dt, \quad (9)$$

where $\boldsymbol{v}$ is the guiding velocity of the particle described by the field (*i.e.*, the function $\Psi$) and $d\boldsymbol{r} = dx\boldsymbol{i} + dy\boldsymbol{j} + dz\boldsymbol{k}$ is a differential element of the position vector, where the $\boldsymbol{i}$, $\boldsymbol{j}$, $\boldsymbol{k}$'s are the usual orthogonal unit vectors along the *x*, *y* and *z*-axis respectively. In the integral (9), the initial conditions are such that at time $t_0$, the particle is at position $\boldsymbol{r} = \boldsymbol{r}_0$. At $(\boldsymbol{r},t) = (\boldsymbol{r}_0, t_0)$, the wavefunction $\Psi$ is such that $\Psi(\boldsymbol{r}_0, t_0) = 0$. This initial condition $\left[ i.e.\ \Psi(\boldsymbol{r}_0, t_0) = 0 \right]$ is required by one of the postulates of QM where the wavefunction must vanish at the end points where the particle is confined[2]. The identity (9) is central to the main result of this reading—it is the guide and lodestar of the present thesis. If the reader has any objections to the final conclusions that we shall arrive at, then they must object to this identity *forthwith*.

Before we proceed, we need to clearly define $\boldsymbol{v}$. Our interpretation of is exactly the same as that handed down to us by the great theoretical physicist—American-British Professor, David Joseph Bohm (1927-1992), in his Pilot Wave Theory which is popularly known as Bohmian Mechanics[3] [8,9]. Remember that all quantum mechanical wavefunctions are functions of the phase $S = \boldsymbol{k} \cdot \boldsymbol{r} \pm \omega t$ *i.e.* $\Psi = \Psi(\boldsymbol{k} \cdot \boldsymbol{r} \pm \omega t)$: where $\boldsymbol{k}$ and $\omega$ are the wave number and angular frequency of the particle. Since according to the *de Broglie's* wave-particle duality hypothesis $\boldsymbol{p} = \hbar \boldsymbol{k}$ and $E = \hbar \omega$ where $\boldsymbol{p}$ and $E$ are the momentum and energy of the particle; it follows that $\Psi = \Psi(\boldsymbol{p} \cdot \boldsymbol{r} \pm Et)$ and $S = \boldsymbol{p} \cdot \boldsymbol{r} \pm Et$. As first noted by *de* Broglie[4] (1927) and latter [8,9] (independently of *de* Broglie's 1927 work), the wavefunction can be written in the form $\Psi = Re^{iS/\hbar}$ where $R = R(\boldsymbol{r},t)$ is a real function which acts as a modulation function of the particle wavepacket. With the wavefunction written in this form, the guiding velocity $\boldsymbol{v}$ of the particle then becomes:

$$\boldsymbol{v} = \left( \frac{\hbar}{m} \right) \text{Im} \left( \frac{\Psi^* \nabla \Psi}{\Psi^* \Psi} \right), \quad (10)$$

where $m$ is the mass of the particle under consideration [8,9, *de* Broglie, 1927]. In (10), the operator "Im" extracts only the imaginary part of a complex function or complex number *i.e.*: if $z = x + iy$ is a complex number, then $\text{Im}(z) = y$ and if $\text{Im}[F(x)] = H(x)$ is a complex valued function where $(G, H)$ are real valued functions of $x$, then $\text{Im}[F(x)] = H(x)$. Thus, for any smooth square integrable and differentiable function $\Psi$ the particle guiding velocity is given by (10).

---

[2]For example, in one dimension, for a particle confined in the region $(0 \leq x \leq a)$, we have $\Psi(0,t) = \Psi(a,t) = 0$.

[3]This theory is also called the *de* Broglie-Bohm theory, the Pilot-Wave model, or the Causal Interpretation of Quantum Mechanics.

[4]*de* Broglie, L., 1928, *in* Solvay 928, p. 119: *Rapports et Discussions du Cinqui ème Conseil de Physique tenu à Bruxelles du* 24 *au* 29 *Octobre* 1927 *sous les Auspices de l'Institut International de Physique Solvay*, Paris: Gauthier-Villars.

Now moving on; the identity (9) can further be simplified. Since $\nabla \cdot \mathbf{v} = 0$, it follows that $\mathbf{v} \cdot \nabla \Psi = \nabla \cdot (\mathbf{v}\Psi)$ because according to the chain rule:

$$\nabla \cdot (\mathbf{v}\Psi) = \mathbf{v} \cdot \nabla \Psi + \Psi \nabla \cdot \mathbf{v} = \mathbf{v} \cdot \nabla \Psi + 0 = \mathbf{v} \cdot \nabla \Psi.$$

With $\mathbf{v} \cdot \nabla \Psi = \nabla \cdot (\mathbf{v}\Psi)$ given, (9) becomes:

$$\Psi \equiv \nabla \cdot \left[ \int_{t_0}^{t} (\mathbf{v}\Psi) \mathrm{d}t \right] = \nabla \cdot \left[ \mathbf{v} \int_{t_0}^{t} \Psi \mathrm{d}t \right]. \tag{11}$$

All we want with the relation (9) is to have the operator "$\nabla$" outside of the integral operator "$\int$". This is important for latter purposes. In the far right of (11), the velocity $\mathbf{v}$ has been removed from inside the integral sign. This is possible because $\mathbf{v}$ has no explicit dependence on time. Any change in $\mathbf{v}$ with respect to time happens not explicitly but implicitly.

Now, as we are going to demonstrate; using the identity (11), one can recast after a few basic algebraic operations, the Schrödinger Equation (1) into a continuity equation. To achieve this, we begin by dividing (1) throughout by "$-i\hbar$", where we will have:

$$\frac{\partial \Psi}{\partial t} = -\frac{i\hbar}{2m} \nabla^2 \Psi + \frac{i}{\hbar} V \Psi. \tag{12}$$

Now, taking the first term on the right hand side of Equation (12), it is clear that:

$$-\frac{i\hbar}{2m} \nabla^2 \Psi = \nabla \cdot \left[ -\frac{i\hbar}{2m} \nabla \Psi \right]. \tag{13}$$

For the second term on the right hand side of (12), using the identity (11) where $V\Psi$ replaces $\Psi$ in this identity i.e. $V\Psi \mapsto \Psi$, it is clear from this identity that:

$$\frac{i}{\hbar} V\Psi = \nabla \cdot \left[ \frac{i\mathbf{v}}{\hbar} \int_{t_0}^{t} V\Psi \mathrm{d}t \right]. \tag{14}$$

Now, adding (13) and (14), we will have:

$$\begin{aligned}\frac{\partial \Psi}{\partial t} &= -\frac{i\hbar}{2m} \nabla^2 \Psi + \frac{i}{\hbar} V\Psi \\ &= -\nabla \cdot \left[ +\frac{i\hbar}{2m} \nabla \Psi - \frac{i\mathbf{v}}{\hbar} \int_{t_0}^{t} V\Psi \mathrm{d}t \right].\end{aligned} \tag{15}$$

This can now be written as:

$$\frac{\partial \Psi}{\partial t} + \nabla \cdot \mathbf{J}_{\Psi}^{S} = 0, \tag{16}$$

where the new current term $\mathbf{J}_{\Psi}^{S}$ is given by:

$$\mathbf{J}_{\Psi}^{S} = \frac{i\hbar}{2m} \nabla \Psi - \frac{i\mathbf{v}}{\hbar} \int_{t_0}^{t} V\Psi \mathrm{d}t. \tag{17}$$

What this means is that there is a corresponding conserved probability amplitude current for the probability current. The quantity $J_{\mu}^{\Psi} = (\Psi, \mathbf{J}_{\Psi}^{S}) : (\mu = 0,1,2,3)$ comprises the four probability amplitude current. Though this is a very trivial result, if it is correct (as we believe) and acceptable, it is a new result in QM.

## 3. Klein-Gordon Probability as Currently Understood

For a free particle of rest mass $m_0$ and wavefunction $\Psi$, the Klein-Gordon Equation describing this particle is given by:

$$\nabla^2 \Psi - \frac{1}{c^2} \frac{\partial^2 \Psi}{\partial t^2} = \left( \frac{m_0 c}{\hbar} \right)^2 \Psi. \tag{18}$$

This equation is named after the physicists Oskar Klein and Walter Gordon, who in 1927 proposed it as an equation describing relativistic Electrons. The Klein-Gordon Equation was first considered as a quantum mechanical wave equation by Schrödinger in his search for an equation describing *de* Broglie waves. In his final presentation in January 1926 where he proposed the Schrödinger Equation, Schrödinger discarded this equation because when the Coulomb potential is in-cooperated for the case of an Electron-Proton system, it did not give the correct predictions for the hydrogen atom as Schrödinger expected.

Now, we would like to develop for the Klein-Gordon Equation the usual expressions for probability density function and the corresponding probability current density similar to the Schrödinger case. This is a task that is considered a bit tricky as compared to the Schrödinger case because the Klein-Gordon Equation is second differential equation. If we take the probability density function $\rho = \Psi^* \Psi$ and then differentiate it with respect to time, what we get is (5), and from this point, if we are to follow the Schrödinger prescription, we should be able to arrive at a continuity equation containing the probability density function and the corresponding probability current density by substituting the time derivatives of the wavefunction.

Now—here lies the problem; most textbooks will tell you that one is not able to proceed to find the continuity equation for the above equation from the Schrödinger prescription simply because the Klein-Gordon Equation does not have a first-order derivative that would enable a straight substitution [see e.g. Ref. 10, pp. 576-578]. So what is typically done is to work backwards, that is start from the known Schrödinger probability current density and proceed from there to see if one can find a corresponding probability density function. As we all know, one does arrive at a continuity equation, namely:

$$\frac{\partial \rho_{KG}}{\partial t} + \nabla \cdot \mathbf{J}_{\rho}^{S} = 0, \tag{19}$$

where the Klein-Gordon probability is given by:

$$\rho_{KG} = \frac{i\hbar}{2m_0 c^2}\left(\Psi^* \frac{\partial \Psi}{\partial t} - \Psi \frac{\partial \Psi^*}{\partial t}\right). \quad (20)$$

There is no need for us to go through the full derivation of the Klein-Gordon probability continuity equation as this can readily be found in most textbooks of quantum mechanics. Further, there is no need to demonstrate that this probability density function leads to negative probabilities for particles of negative energy as this is well anchored in most quantum mechanics textbooks. What we shall do is to point out that there is a loophole in this derivation and this loophole is deeply embedded in the fact that:

*The Klein-Gordon Equation is not a first order differential equation which would allow for a smooth and straight forward substitution of the time derivative of the magnitude of the Klein-Gordon wavefunction into Equation* (4) *directly from the Klein-Gordon Equation, so as to derive the probability continuity equation; because of this, one has to seek other alternative means.*

Why not force the Klein-Gordon Equation to produce a probability current density under these conditions? The legitimate rules of mathematics allow for this, why not go for it?! This is our borne of contention.

It is perhaps important that we mention here that in 1934, two other distinguished Austrians, Wolfgang Pauli (1904-1982) and Victor Frederick Weisskopf (1908-2002) discovered what is hailed as a suitable interpretation of the Klein-Gordon Equation within the scope of quantum field theory. Treating it [Klein-Gordon wavefunction] like a field equation analogous to Maxwell's Equations for an electromagnetic field, they quantized it, so that $\Psi$ became an operator [11]. This made the Klein-Gordon theory more acceptable and since then, there appears to have been some acceptable and believable order in the Klein-Gordon world.

As will be seen latter in § (4), if the main reason for adding the Klein-Gordon probability density function that leads to negative probabilities is that it emerges from the continuity equation constructed out of the second order differential Klein-Gordon Equation; then, this way of arriving at the probability density function can be challenged as there is another way to arrive at a continuity equation from the Klein-Gordon Equation, this equation involves the magnitude of the Klein-Gordon wavefunction. After all, no one has made a direct measurement to test the correctness or lack thereof the Klein-Gordon probability and the Klein-Gordon probability current density, it is just but an "agreed upon interpretation".

**Klein-Gordon Probability Amplitude Current**

First things first, we need to categorically state one thing which is clear to all: which is that, the Klein-Gordon Equation is not cast in stone; the meaning of which is that it can be rewritten in different but mathematically equivalent forms provided one applies permissible and legitimate mathematical operations to it. So, just as we have done in the Schrödinger case, we will write down the continuity equation of the Klein-Gordon Equation which involves the probability amplitude $(\Psi)$ and not the probability density function $(\rho)$. By integrating the Klein-Gordon Equation with respect to time throughout, one consequently recasts this equation into the form:

$$\frac{1}{c^2}\frac{\partial \Psi}{\partial t} - \nabla \cdot \left[\nabla\left(\int \Psi dt\right)\right] = -\left(\frac{m_0 c}{\hbar}\right)^2 \int \Psi dt. \quad (21)$$

In the above equation—for our convenience vis-a-vis the labour of typing, we have not put the limits to the integral as we have done in (11); from here and after, we shall no-longer put the limits. It shall be assumed that the reader knows them. That said, written in this form *i.e.* the Klein-Gordon Equation in the form (21), one can easily construct a continuity equation. The strategy here being championed is to rewrite the second order Klein-Gordon differential equation as a first order differential equation in terms of time as is the case with the Schrödinger Equation. Having done this; using the fact stated in (11), it is not difficult to deduce that (21) can further be written as:

$$\frac{\partial \Psi}{\partial t} = \nabla \cdot \left[\nabla\left(\int c^2 \Psi dt\right) - \left(\frac{m_0 c^2}{\hbar}\right)^2 \left(\iint \Psi dt dt\right)v\right]. \quad (22)$$

Thus setting:

$$\boldsymbol{J}_\Psi^{KG} = \nabla\left(\int c^2 \Psi dt\right) - \left(\frac{m_0 c^2}{\hbar}\right)^2 \left(\iint \Psi dt dt\right)v, \quad (23)$$

it is easy to see that:

$$\frac{\partial \Psi}{\partial t} + \nabla \cdot \boldsymbol{J}_\Psi^{KG} = 0. \quad (24)$$

This is our desired equation. What this means is that the probability amplitude has a corresponding current. In the language of Einstein's Special Theory of Relativity (STR), it means we can talk of a four probability amplitude comprising the probability amplitude and the probability amplitude current *i.e.* $\left(\rho, \boldsymbol{J}_\Psi^{KG}\right)$.

## 4. New Klein-Gordon Probability Continuity Equation

Now, we come to the main theme of this reading. In (21) we have written the Klein-Gordon Equation with the time derivative to first order. For convenience, we shall rewrite this Equation (21) here as:

$$\frac{\partial \Psi}{\partial t} = \nabla \cdot \left[\nabla\left(\int c^2 \Psi dt\right)\right] - \left(\frac{m_0 c^2}{\hbar}\right)^2 \int \Psi dt. \quad (25)$$

Now, multiplying this equation throughout by the complex conjugate of the wavefunction, that is:

$$\Psi^* \frac{\partial \Psi}{\partial t} = \Psi^* \nabla \cdot \left[ \nabla \left( \int c^2 \Psi dt \right) \right] - \left( \frac{m_0 c^2}{\hbar} \right)^2 \Psi^* \int \Psi dt, \quad (26)$$

and then taking the complex conjugate of this same equation and then multiplying it by the wavefunction, one arrives at:

$$\Psi \frac{\partial \Psi^*}{\partial t} = \Psi \nabla \cdot \left[ \nabla \left( \int c^2 \Psi^* dt \right) \right] - \left( \frac{m_0 c^2}{\hbar} \right)^2 \Psi \int \Psi^* dt. \quad (27)$$

Adding these two Equations (26) and (27), one obtains:

$$\Psi^* \frac{\partial \Psi}{\partial t} + \Psi \frac{\partial \Psi^*}{\partial t} = c^2 \left( \Psi^* \int \nabla^2 \Psi + \Psi \int \nabla^2 \Psi^* \right) dt - \left( \frac{m_0 c^2}{\hbar} \right)^2 \left( \Psi^* \int \Psi + \Psi \int \Psi^* \right) dt. \quad (28)$$

The left hand side is obviously equal to the time derivative of the probably density function $\rho = \Psi^* \Psi$ i.e. $d\rho/dt = d(\Psi^*\Psi)/dt$. To simplify the right hand side, we have to make use of the identity in Equation (11); doing so, we will have:

$$c^2 \left( \Psi^* \int \nabla^2 \Psi + \Psi \int \nabla^2 \Psi^* \right) dt = -\nabla \cdot \left[ -\mathbf{v} \int \left\{ c^2 \left( \Psi^* \int \nabla^2 \Psi + \Psi \int \nabla^2 \Psi^* \right) dt \right\} dt \right], \quad (29)$$

and upon further reduction, the above becomes:

$$-\left( \frac{m_0 c^2}{\hbar} \right)^2 \left( \Psi^* \int \Psi + \Psi \int \Psi^* \right) dt = -\nabla \cdot \left[ \mathbf{v} \int \left\{ \left( \frac{m_0 c^2}{\hbar} \right)^2 \left( \Psi^* \int \Psi + \Psi \int \Psi^* \right) dt \right\} dt \right]. \quad (30)$$

Now, adding (29) and (30), we obtain, $-\nabla \cdot \mathbf{J}^{KG}$, where $\mathbf{J}^{KG}$ is the new Klein-Gordon probability current density such that $\mathbf{J}^{KG} = \rho \mathbf{v}$ where $\rho$ is what we shall call the Klein-Gordon probability charge density. This charge density is such that:

$$\rho = -\int \left[ \int \left\{ c^2 \left( \Psi^* \int \nabla^2 \Psi + \Psi \int \nabla^2 \Psi^* \right) dt \right\} - \left\{ \left( \frac{m_0 c^2}{\hbar} \right)^2 \left( \Psi^* \int \Psi + \Psi \int \Psi^* \right) dt \right\} \right] dt. \quad (31)$$

This new Klein-Gordon probability current density $\mathbf{J}^{KG} = \rho \mathbf{v}$ is what leads us to positive definite probabilities[5]. All our seemingly naïve efforts lead us to recast the Klein-Gordon Equation into the new continuity equation:

$$\frac{\partial \rho}{\partial t} + \nabla \cdot \mathbf{J}_\rho^{KG} = 0. \quad (32)$$

In this manner, just as in the Schrödinger case, the magnitude $(\rho = \Psi^*\Psi \geq 0)$ of the wavefunction gives a positive definite probability and is part of a four current. The quantity $(\rho, \mathbf{J}_\rho^{KG})$ is the new four Klein-Gordon probability current that leads to positive definite probabilities *via* the new definition of the Klein-Gordon probability current $\mathbf{J}_\rho^{KG}$. We feel and believe this approach is the correct approach to understanding the Klein-Gordon Equation. It contains no negative probabilities but real and objective probabilities just as in the Schrödinger case. It is our modest and vested opinion that it is much easier to try and understand $\mathbf{J}_\rho^{KG}$ as the new Klein-Gordon probability current density, than to try and justify negative probabilities as Dirac, Feynman and many others have attempted (with great pain), only to fail.

## 5. Discussion and Conclusion

### 5.1. General Discussion

At this juncture, we are of the candid view that the reader will concur with us that—from the rather trivial presentation made herein; the existence of negative quantum mechanical probabilities depends on the choice of the probability current density that one has made. In moving from Schrödinger's theory (which officially was first to be discovered) to the Klein-Gordon theory, it is the probability current density that is held sacrosanct [e.g. Ref. 10, pp. 576-578], the meaning or suggestion of which is that it must be the important quantity, otherwise there really would be no reason to preserve it. We ask our dear reader; could there be any other better reason to do so besides that it is a "sacrosanct quantity"? We have suggested otherwise, that it is the probability density function in the Schrödinger theory that must be held sacrosanct when we move over to the Klein-Gordon theory. This way of looking at the Klein-Gordon theory resolves (once and for all-time) the negative probabilities that bedevil and bewilder this theory. Because we here have shown that these negative probabilities can be gotten rid of completely, we believe this reading is a significant contribution to physics insofar as the endeavours to deciphering and fathoming the meaning of Klein-Gordon's negative probabilities is concerned.

---

[5]In comparison—the expression for the Schrödindger probability current density 'is more beautiful' than the new expression for the Klein-Gordon probability current. In-accordance with Weyl's doctrine, our work is always to unite the *Truth* with the *Beautiful*. In all honesty, there is no derivable Truth in negative probabilities of Klein and Gordon, thus how can we unite these together into an everlasting ornament? There is obviously an element of truth in the new positive definite Klein-Gordon probability. Somehow, there must exist a hidden beauty in the new Klein-Gordon probability current.

In all history of human experience, it is important to note that no single experiment has been performed to date to directly measure the probability and the probability current implied by the quantum mechanical wavefunction. For example, the wavefunction of the hydrogen atom as deduced from the Schrödinger Equation is known yet no one has measured directly that the Electron in the hydrogen atom is found at the position that it is expected to be with the predicted frequency. The probability interpretation is an interpretation that strongly appears to work very well, especially when dealing with ensembles. What this means is that the currently accepted Klein-Gordon probability can be revised as we have done. If what is required of this probability is that it satisfies the continuity equation where this probability density function has a corresponding probability current density —well then—we have shown that there exists such a positive definite probability satisfying the continuity equation. It is difficult to dismiss the present approach (proposal). Whether or not mainstream science will seriously consider our approach (proposal made herein), only time will tell.

If our suggestion is correct and acceptable (as we believe it to be), then, one is lead to wonder what trajectory physics might have taken if what we have just presented were known to Dirac and his contemporaries. This is so especially given that Dirac was largely motivated by the desire to get reed of the negative probabilities that appear in the Klein-Gordon theory. There is nothing exotic or new about the ideas that we have presented, it is just a different way of looking at things. The only thing that appears to make this approach of importance, is that it allows us to settle once and for all-time the nagging problem of negative probabilities.

Clearly, if the present presentation was available and acceptable to Dirac before the advent of his equation; then, if he [Dirac] was to discover the Dirac Equation as he did, he would have arrived at it from a different point of departure altogether. I wonder what his motivation would have been. Trying to imagine what his point of departure would have been, leads me into the wilderness of thought—analogous to chasing after the rainbow, the wind or one's own shadow. Perhaps, out of mathematical curiosity, beauty and elegance, he simply would have sought for an equation linear in both the time and space derivatives.

On the other hand, since negative probabilities are intrinsically tied to negative energy-mass particles, the non-existence of negative probabilities would mean that the existence of negative energy-mass particles has no problem in principle. This invariably means that negative energy-mass particles must be considered without any prejudice whatsoever as they have equal legitimacy to exist. Our only concern would be what these negative energy-mass particles are; are they Dirac's antimatter, or Dirac's sea of invisible energy-mass particles? Current thinking due to Richard Feynman is that negative energy-mass particles are antiparticles. These antiparticles have positive energy and the reason for this is that they are thought to be negative energy-mass particles moving back in time, in which case they would appear to have positive energy-mass [see e.g. Ref. 12, pp. 66-70]. The perfect symmetry of Dirac's theory allows a negative energy-mass particle that is moving forward in time to look identical to a positive energy-mass particle moving back in time.

## 5.2. Conclusions

Assuming the correctness (or acceptableness) of the ideas presented herein, we hereby make the following conclusion:

1) This reading has demonstrated that negative probabilities can be avoided in quantum mechanics by making an appropriate choice of the Klein-Gordon probability current.

2) By avoiding the negative probabilities as suggested herein, it is seen that negative energy particles will still exist with the important difference that they are no longer attached to the meaningless negative probabilities as is the case in the original Klein-Gordon theory. These negative energy particles have a real positive definite probability of existence.

## 6. Acknowledgements

I am grateful to the four anonymous Reviewers for their effort that greatly improved and refined the arguments presented herein. Further, I am grateful to the National University of Science and Technology's Research & Innovation Department and Research Board for their unremitting support rendered toward my research endeavours; of particular mention, Dr. P. Makoni and Prof. Y. S. Naik's unwavering support. This publication proudly acknowledges a GRANT from the National University of Science and Technology's Research Board.